# Mechanical properties of carbon nanotube reinforced polymer nanocomposites: a coarse-grained model

Behrouz Arash[1,*], Harold S. Park[2], Timon Rabczuk[1]

[1]*Institute of Structural Mechanics, Bauhaus Universität-Weimar, Marienstr 15, D-99423 Weimar, Germany*

[2]*Department of Mechanical Engineering, Boston University, Boston, Massachusetts 02215, USA*

**Abstract**

In this work, a coarse-grained (CG) model of carbon nanotube (CNT) reinforced polymer matrix composites is developed. A distinguishing feature of the CG model is the ability to capture interactions between polymer chains and nanotubes. The CG potentials for nanotubes and polymer chains are calibrated using the strain energy conservation between CG models and full atomistic systems. The applicability and efficiency of the CG model in predicting the elastic properties of CNT/polymer composites are evaluated through verification processes with molecular simulations. The simulation results reveal that the CG model is able to estimate the mechanical properties of the nanocomposites with high accuracy and low computational cost. The effect of the volume fraction of CNT reinforcements on the Young's modulus of the nanocomposites is investigated. The application of the method in the modeling of large unit cells with randomly distributed CNT reinforcements is examined. The established CG model will enable the simulation of reinforced polymer matrix composites across a wide range of length scales from nano to mesoscale.

**1. Introduction**

Since their discovery, carbon nanotubes (CNTs) have attracted intense attention among scientists in various disciplines due to their outstanding electrical, mechanical and thermal properties [1, 2]. The remarkable material properties have made CNTs as excellent candidates in a wide range of applications such as nano-sensors and atomic transportation [3, 4]. In addition, the ultra-high Young's modulus and tensile strength of nanotubes [5] are promising ultra-high-strength reinforcements in high-performance polymer matrix composites [6].

In the past decade, extensive experimental and theoretical studies [7-11] have been conducted on the mechanical properties of reinforced polymer composites to facilitate the development of the materials. Although a number of experimental investigations have been carried out in the literature to determine the mechanical characteristics of the nanocomposites, they provide insufficient insight into molecular scale processes such as the interfacial interactions between nanotubes and matrix. Furthermore, experimental efforts encounter difficulties in fabricating short-fiber polymer composites reinforced by uniformly distributed CNTs with desired sizes. The main reason for these drawbacks is attributed to the limited resolution of experimental techniques at the nanoscale. Molecular dynamics (MD) simulations, however, present alternative methods to generate detailed information of atomistic mechanisms such as stick–slip mechanisms at the nanoscale, the stress-strain behavior and the interfacial interactions between matrix and reinforcements [8, 10, 12-14]. MD simulations also enable the interpretation of experimental data, and further open a route to new designs of nanocomposites. Hence, molecular simulations are indispensable in understanding mechanical properties of CNT/polymer composites. Frankland et al. [15] investigated stress–strain curves of (10, 10) single-walled CNTs (SWCNTs) reinforced polyethylene matrix composites. Based on their simulation results, long SWCNT fibers with effectively infinite lengths show a significant increase in the stiffness of the nanocomposites, while no considerable enhancement is observed for short nanotube reinforcements with a length-to-diameter aspect ratio of 4. Zhu et al. [16] studied the mechanical properties of an epoxy Epon 862 matrix with a size of $4.028 \times 4.028 \times 6.109$ nm$^3$ reinforced by short and long (10, 10) SWCNTs. The short and long CNTs have a length-to-diameter aspect ratio of 2.15 and 4.5, respectively. They reported that short CNT fibers can increase the Young's modulus of the polymer matrix up to 20%, while long nanotubes with effectively infinite length significantly increase the stiffness of the representative volume element (RVE) more than 10 times compared to a pure Epon 862 matrix. Han and Elliott [17] studied the elastic properties of poly (methyl methacrylate) (PMMA) matrix reinforced by a (10, 10) SWCNT with an effectively infinite length. Based on their simulation results, the longitudinal Young's modulus of a CNT/PMMA RVE with nanotube volume fractions of 12 and 17 % are 94.6 and 138.9 GPa, respectively. Mokashi et al. [18] presented molecular simulation studies on the elastic moduli and tensile strengths of an amorphous polyethylene polymer matrix reinforced by a (10, 10) CNT. The simulation cell size for the nanocomposite was set to be $4.5 \times 4.5 \times 4.4$ nm$^3$. Their simulation results show that the longitudinal Young's modulus of the CNT/polymer composite with a nanotube volume fraction of 11.25 % is obtained to be 82 GPa, which is about 25 times of that for pure amorphous polyethylene. Molecular simulation studies on the mechanical properties of short CNT reinforced Poly (vinylidene fluoride) (PVDF) matrix composites [19] indicated that an introduction of a (5, 5) SWCNT with a length of 2 nm can increase the Young's modulus of a CNT/PVDF unit cell by 1 GPa. The simulation unit cell consists of the (5, 5) SWCNT embedded in 60 PVDF chains, where the nanotube volume fraction is 1.6 %. The mechanical behavior of CNT/PMMA matrix composites under tension was investigated by Arash et al. [8]. They proposed a new method for evaluating the elastic properties of the interfacial region of CNT/polymer composites. Their simulation results on the elastic properties of a PMMA polymer matrix with a size of 3.7×3.7×8 nm$^3$ reinforced by a (5, 5) CNT reveal that the Young's modulus of the composite increases from 3.9 to 6.85 GPa with an increase in the length-to-diameter aspect ratio of the CNT from 7.23 to 22.05.

---
[*] Author to whom correspondence should be addressed. E-mail address: behrouz.arash@uni-weimar.de, Tel: +49 3643 584511.

Although molecular simulations have been widely used in modeling nanocomposites, the huge computational effort required by the simulations severely limits their applicability to small molecular systems over a limited time scale. In order to overcome these drawbacks, coarse-grained (CG) models beyond the capacity of molecular simulations have been developed in the literature [20-22]. The principle of CG models is to map a set of atoms to a CG bead, which enables to extend the accessible time and length-scales while maintaining the molecular details of an atomistic system. The main challenge is therefore to develop a CG model that reproduces the same physical behavior as the atomistic reference system. Up to now, many CG models have been developed for polymer materials in the literature [21, 23, 24]. Recently, the application of these approaches in modeling graphenes and CNTs has been also examined [25-28]. Cranford et al. [26] proposed a CG model derived directly from atomistic simulations by enforcing the assertion of energy conservation between atomistic and mesoscopic models. The CG model was utilized to study the folding of graphenes at the mesoscopic length-scale. A CG model of graphene was established by Ruiz et al. [27] using a strain energy conservation approach. The model was shown to be able to accurately predict the elastic and fracture behavior of graphenes with a ~200 fold increase in computational speed compared to atomistic simulations. Zhao et al. [28] parameterized the CG stretching, bending and torsion potentials of SWCNTs to study their static and dynamic behaviors. They also calibrated the non-bonded CG potential between CG beads using the van der Waals (vdW) cohesive energy between SWCNTs in a bundle derived from analytical approaches. The CG model was shown to have great potential in the analysis of the mechanical properties of CNT bundles and buckypapers with low computational cost compared to atomistic simulations. Although CG models have been extensively employed for studying the material properties of polymer materials, graphenes and CNTs, a more comprehensive CG model of CNT/polymer composites concerning interactions between polymer chains and nanotubes is still necessary. The CG model, proving the underlying physics of interactions between nanotube reinforcements and a polymer matrix, enables to study the material properties of the nanocomposites across a wide range of length and time scales.

This study aims to develop a CG model of CNT/polymer composites, with a particular emphasis on capturing the non-bonded interactions between nanotube reinforcements and polymer chains. The parameters of the CG potentials are derived based on the strain energy conservation between CG models and full atomistic systems. The applicability of the CG model in reproducing the elastic response of the nanocomposites is examined using results obtained from molecular simulations. The effect of the nanotube volume fraction on the Young's modulus of the composites is studied. The application of the model in predicting the elastic properties of large polymer unit cells with randomly distributed CNT reinforcements is studied.

**2. Coarse-grained model**

The applicability of a CG model depends on the compatibility between CG and atomistic force fields. There are three major approaches to construct CG force fields: the iterative Boltzmann (IB) inversion method [21, 29, 30], the Martini force field [31, 32] and methods based on the strain energy conservation between a full atomistic system and its corresponding CG model [26-28]. According to the Henderson theorem [33] the IB method should give unique two body CG interactions. However, the convergence is difficult to achieve in practice [34]. Although the Martini force field has been parameterized to simulate lipid, proteins and surfactant systems [31, 35], it requires recalibration for describing the elastic properties of CNTs. In addition, the Martini force field does not provide structural details for a specific system. Methods based on the strain energy conservation, in contrast, have been successfully applied to reproduce the elastic response of CNTs, and to capture their structural details [28]. The principle of the approaches is based on calibrating the CG potentials by enforcing the assertion of energy conservation between CG and molecular models. As a result, the mechanical properties of a nanostructure, such as the Young's modulus and the adhesion energy per surface area, predicted by the CG model are in fair agreement with those obtained from molecular simulations.

Herein, we develop a CG model of CNTs and PMMA polymer chains, which enables simulating nanotube reinforced polymer composites. The force fields of the CG model are decomposed into bonded and non-bonded potential functions. The total potential energy, $E_{total}$, of a system is therefore written as the sum of energy terms associated with the variation of the bond length, $E_b$, the bond angle, $E_a$, the dihedral angle, $E_d$, the vdW interactions, $E_{vdW}$, and the constant free energy of the system, $U_0$, as $U_{total} = \sum_i E_{b_i} + \sum_j E_{a_j} + \sum_k E_{d_k} + \sum_{lm} E_{vdW_{lm}} + U_0$. The functional forms of the contributing terms for a single interaction are as follows:

$$E_b(d) = \frac{k_d}{2}(d - d_0)^2, \tag{1a}$$

$$E_a(\theta) = \frac{k_\theta}{2}(\theta - \theta_0)^2, \tag{1b}$$

$$E_d(\phi) = \frac{k_\phi}{2}[1 + \cos 2\phi], \tag{1c}$$

$$E_{vdW}(r) = D_0\left[\left(\frac{r_0}{r}\right)^{12} - 2\left(\frac{r_0}{r}\right)^6\right], \tag{1d}$$

where $k_d$ and $d_0$ are the spring constant of the bond length and the equilibrium bond distance, respectively; $k_\theta$ and $\theta_0$ are respectively the spring constant of the bond angle and the equilibrium bond angle; $k_\phi$ and $\phi$ are the spring constant of the dihedral angle and the dihedral angle, respectively. $D_0$ and $r_0$ are the Lennard-Jones parameters associated with the equilibrium well depth and the equilibrium distance, respectively.

In the following section, the parameters of the functional forms are calibrated for PMMA polymer and CNTs through verification processes with molecular simulations. In the molecular simulations, COMPASS force field [36] is used to describe intermolecular interactions. It is the first ab initio force-field that enables an accurate prediction of the mechanical behavior of CNTs and polymers. The non-bonded interactions are modeled using the vdW and coulombic interaction energy terms. A potential cutoff of 1.2 nm is used in calculation of the non-bonded interactions. Partial charges of atoms are also assigned using Qeq method [37]. The conjugate-gradient energy minimization method [38] is implemented to achieve the minimum energy configuration of a full atomistic system.

*2.1. Coarse-grained stretching potentials*

The first step of a coarse-graining procedure is to map a set of atoms of a full atomistic system onto a CG bead for simplifying the structure. The present CG model for PMMA polymer chains is developed based on using one bead per methyl methacrylate ($C_5O_2H_8$) monomer. The center of the CG bead is chosen to be the center of mass of the monomer (See Figs. 1 (a) and (b)). The pseudoatom with an atomic mass of 100.12 amu and a radius of 2.15 Å is defined as P bead. The CG model of the polymers enables a 15 fold decrease in the number of degrees of freedom (DOF) of the system compared to a full atomistic system.

For small deformations along the centers of a two-monomer system illustrated in Fig. 1 (a), the total potential energy, $U$, obtained by molecular simulations can be equalized to the stretching potential of a two-bead CG model shown in Fig. 1 (b) as $U = \frac{k_d}{2}(d - d_0)^2 + U_0$. $d$ is the distance between the two monomers, and $d_0$ is the equilibrium distance measured to be 4.02 Å from molecular simulations. The spring constant of the bond length is then given by the second derivative of the potential energy with respect to the bond length as $k_d = \frac{\partial^2 U}{\partial d^2}$. Fig. 1 (c) presents the variation of potential energy of the two-monomer system versus deformations from which the spring constant of the bong length is calculated to be $k_d = 194.61\ kcal/mol/Å^2$.

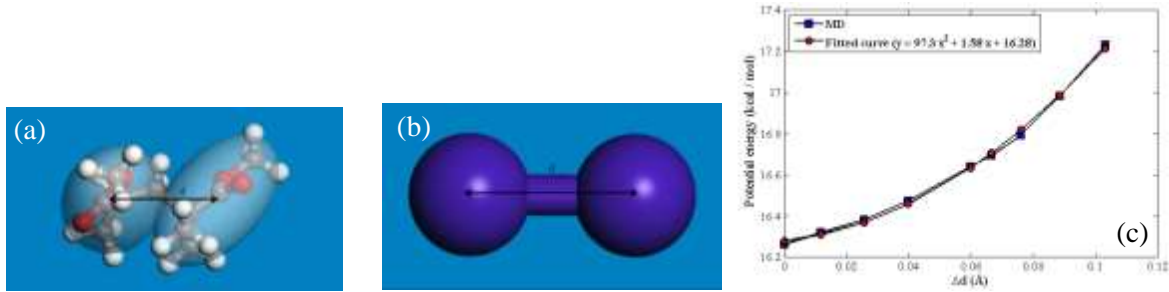

Fig. 1. (a) Two monomers of a PMMA polymer chain, (b) CG model of the monomers made of two P beads, and (c) the variation of potential energy versus deformation using molecular simulations.

The CG model is further developed for modeling (5, 5) CNTs in which each bead represents five atomic rings. The atomic mass and radius of the bead (defined as C bead) are set to be 600.55 amu and 3.96 Å, respectively. The CG model of CNTs decreases the number of DOF of the system 50 fold with respect to full atomistic simulations. Figs. 2 (a) and (b) show a (5, 5) CNT with ten atomic rings and its two-bead CG counterpart, respectively. Similar to previous simulations, the spring constant of the bond length is determined for the defined beads. The potential energy of the CNT under a longitudinal deformation obtained by molecular simulations is equated to the potential energy of the CG model. Fig. 2(c) presents the variation of potential energy of the (5, 5) CNT under a longitudinal deformation by which the spring constant of the bond length of the CG model illustrated in Fig. 2(b) is calculated to be $k_d = 1610.24\ kcal/mol/Å^2$. The equilibrium distance is also measured to be $d_0 = 6.03$ Å.

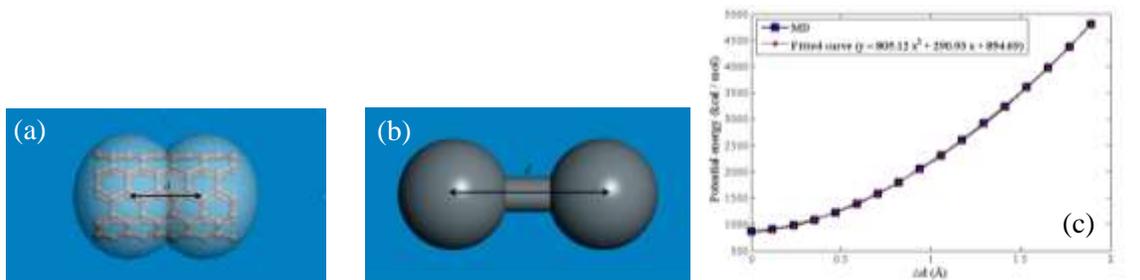

Fig. 2. (a) A (5, 5) CNT with 10 rings of carbon atoms, where each 5 rings is modeled as a bead in coarse-grained model, (b) CG model of the CNT made of two C beads, and (c) the variation of potential energy versus deformation using molecular simulations.

## 2.2. Coarse-grained bending potentials

In the CG model, a three-monomer system with a bond angle, $\theta$, as illustrated in Fig. 3 (a) is simulated with P beads as shown in Fig. 3 (b). The total potential energy of the three monomers, $U$, under a pure bending deformation obtained by MD simulations can be expressed as $U = \frac{k_\theta}{2}(\theta - \theta_0)^2 + U_0$ in the CG system. $\theta_0$ is the equilibrium bond angle, which is measured to be 89.6° from molecular simulations. The spring constant of the bond angle is then obtained from the second derivative of the potential energy with respect to the bond angle as $k_\theta = \frac{\partial^2 U}{\partial \theta^2}$. Fig. 3 (c) presents the variation of potential energy of the three-monomer system versus the bond angle from which the spring constant is calculated to be $k_\theta = 794.89$ kcal/mol/$rad^2$.

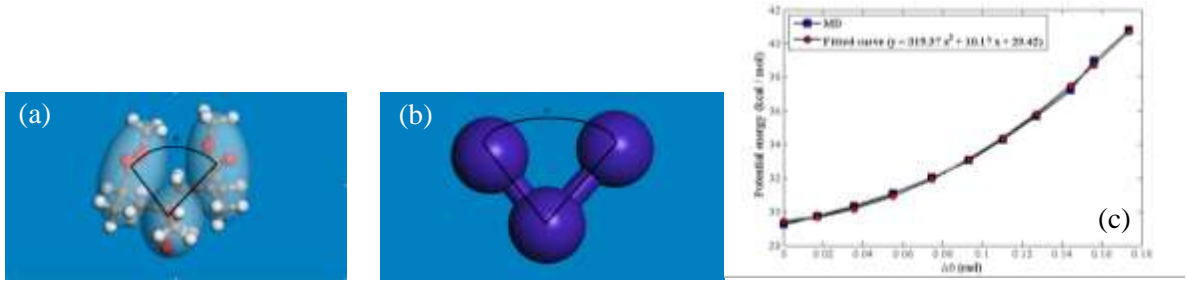

Fig. 3. (a) Three monomers of a PMMA polymer chain, (b) the CG model of the three-monomer system made of three P beads, and (c) the variation of potential energy versus the bond angle ($\theta$) using molecular simulations.

Similarly, the spring constant of the bond angle is determined for C beads. Fig. 4 (a) and (b) show a (5, 5) CNT with fifteen atomic rings and its CG model made of three C beads, respectively. Fig. 3 (c) presents the variation of potential energy of the CNT subjected to a pure bending deformation by which the spring constant of the bond angle for C beads is obtained to be $k_\theta = 66148.01 \: kcal/mol/rad^2$. The equilibrium bond angle for C beads is also obtained to be $\theta_0 = 180°$.

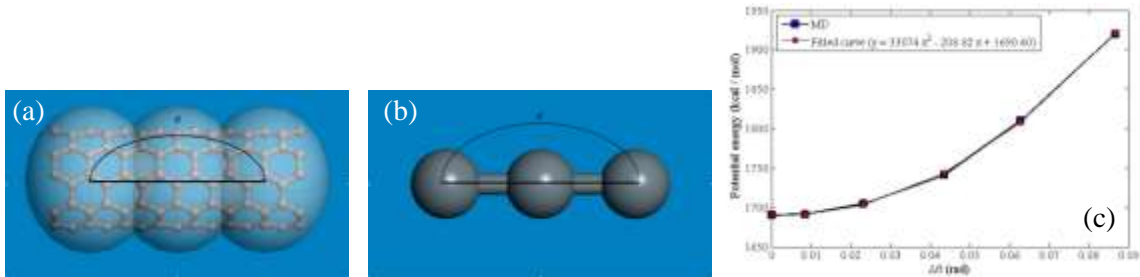

Fig. 4. (a) A (5 ,5) CNT with fifteen atomic rings, (b) the CG model of the nanotube made of three C beads, and (d) the variation of potential energy versus bond angle of the CNT using molecular simulations.

*2.3. Coarse-grained torsion potentials*

In this section, we calibrate the spring constant of the dihedral angle, $k_\phi$, for P and C beads. A four-monomer system and the dihedral angle, $\phi$, between their mean planes are illustrated in Fig. 5 (a). The CG model of the atomistic system made of four P beads is also demonstrated in Fig. 5 (b). In order to determine the spring constant of the dihedral angle of the CG model, the total potential energy of the full atomistic system, $U$, under a pure torsional deformation is equated to the CG force field as $U = \frac{k_\phi}{2}[1 + \cos 2\phi] + U_0$. Next, the least square method is implemented to estimate the optimal magnitude of $k_\phi$ for the best fit between results of MD simulations and the CG model. Fig. 5 (c) presents the variation of potential energy the four-monomer system subjected to a pure torsional deformation obtained by MD simulations and the CG model from which the spring constant of the dihedral angle is calculated to be $k_\phi = 42.05$ kcal/mol. Similarly, the spring constant of the dihedral angle is determined to be 14858.80 kcal/mol for C beads.

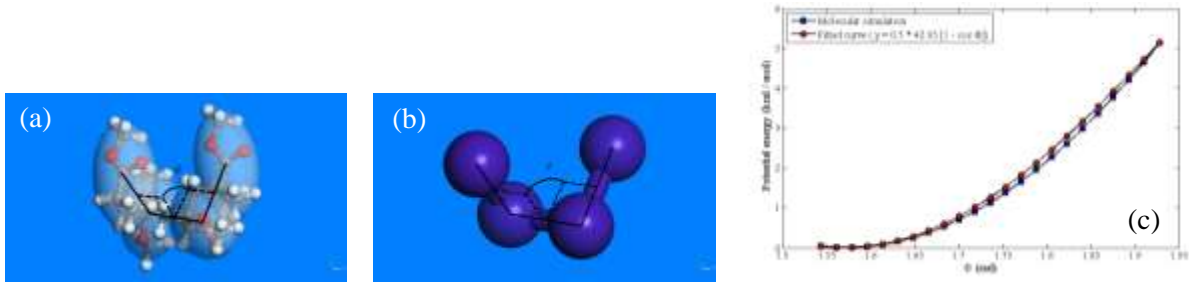

Fig. 5. (a) Four monomers of a PMMA polymer chain, (b) the CG model of the four-monomer system made of four P beads, and (c) the variation of potential energy versus the dihedral angle ($\phi$) using molecular simulations.

*2.4. Coarse-grained van der Waals potentials*

The vdW interactions between CNTs and polymer chains significantly influence the mechanical properties of nanotube/polymer composites. Herein, we discuss how to determine the Lennard-Jones parameters, i.e. the equilibrium well depth ($D_0$) and the equilibrium distance ($r_0$), for the coarse model. Figs. 6 (a) and (b) show two methyl methacrylate monomers and their CG model comprised of two P beads at a distance of $h$, respectively. The distance is measured between the centers of the monomers. Fig. 6 (c) shows the variation of the cohesive energy between the monomers versus their distance obtained by molecular simulations. The cohesive energy is the difference between the total vdW energy of the atomistic system and the vdW energy of each isolated monomer, which can be taken as the vdW interaction between two P beads in the CG model. Therefore, the variation of the energy is utilized to adjust the Lennard-Jones parameters in Eq. 1 (d). The minimum value of the cohesive energy is chosen to be the equilibrium well depth as $D_0 = 1.34\ kcal/mol$, and the distance at which the minimum energy occurs is set to be the equilibrium distance as $r_0 = 6.53$ Å. The variation of vdW energy between the two monomers presented in Fig. 6 (c) confirms that there is good agreement between results predicted by molecular simulations and those obtained from the CG model. Likewise, the equilibrium well depth and the equilibrium distance of the vdW interaction between two C beads are attained to be $D_0 = 10.68\ kcal/mol$ and $r_0 = 9.45$ Å, respectively. The Lennard-Jones parameters calculated for the vdW interaction between a C and a P bead is further calculated to be $D_0 = 2.80\ kcal/mol$ and $r_0 = 7.71$ Å.

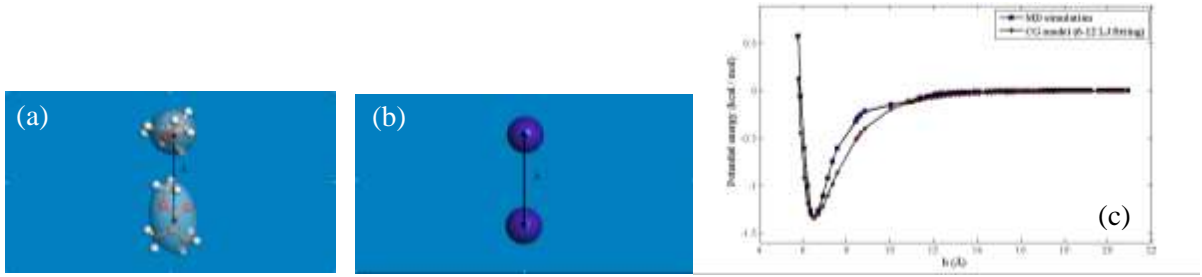

Fig. 6. (a) Two monomers of PMMA polymer, (b) coarse-grained model of the monomers, and (c) coarse-grained model of the monomers, and (c) the variation of vdW energy versus distance between the two monomers using molecular simulations.

## 3. Mechanical properties of nanotube-reinforced composites

### 3.1. Mechanical behavior of polymer chains and CNTs

In order to investigate the mechanical properties of CNT/PMMA composites, the validation of the calibrated CG force field parameters is first explored. For this, the mechanical behavior of PMMA polymers and CNTs under axial tensile deformation predicted by the CG model is verified by molecular simulation results. Fig. 7 (a) illustrates a PMMA polymer chain with 30 monomers and its CG model made of 30 P beads. The initial length of the chain is about 66.10 Å and a static tensile deformation with a strain rate of 0.1 % is applied to the chain. The variation of potential energy of the chain subjected to the deformation obtained by molecular simulations and the CG model is presented in Fig. 7 (b). The equivalent spring constant of the polymer chain can be calculated from the second derivative of the potential energy with respect the axial deformation. From Fig. 7 (b), the equivalent sprint constant predicted by molecular simulations and the CG model is respectively calculated to be 2.005 and 1.94 $kcal/mol\ /Å^2$, revealing a percentage difference less than 3.3%. The simulation results justify the applicability of the CG force field parameters derived for PMMA polymer chains. Likewise, the applicability of force field parameters calibrated for C beads in the analysis of CNTs is examined. Fig. 7 (c) shows a (5, 5) CNT with periodic boundary conditions.

The CG model of the nanotube made of C beads is also demonstrated in Fig. 7 (d). Fig 7 (e) presents the variation of potential energy of the CNT subjected to axial tensile deformation obtained by molecular simulations and the CG model. The static tensile deformation is applied with a strain rate of 0.1 %. From Fig. 7 (e), the equivalent spring constant of the nanotube obtained from molecular simulations and the CG model is 97.64 and 97.24 $kcal/mol/Å^2$, respectively.

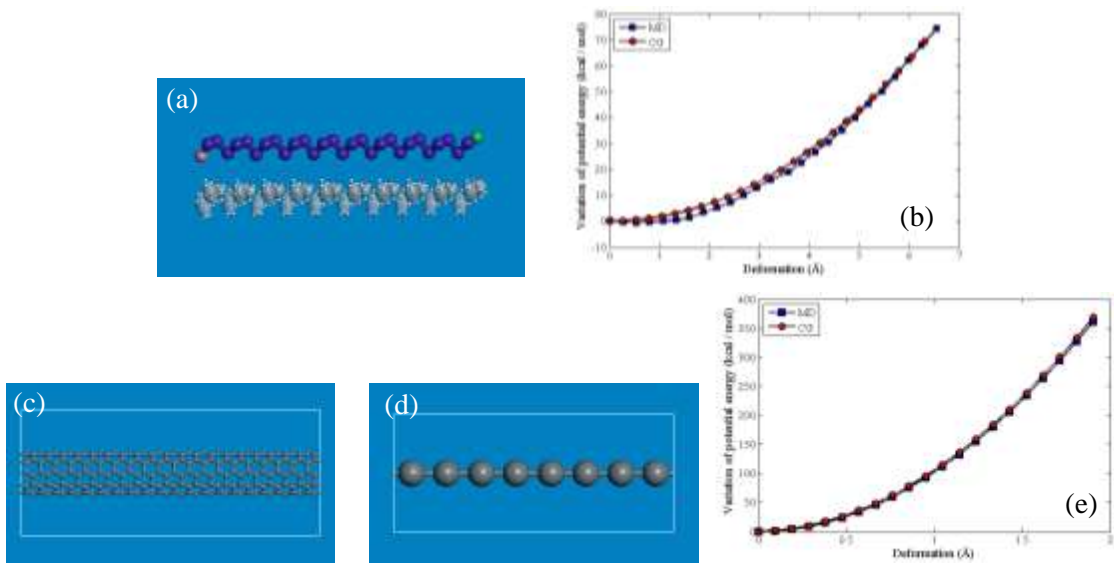

Fig. 7. (a) A PMMA polymer chain made of 30 monomers and its CG model made of 30 P beads, (b) the variation of potential energy versus tensile longitudinal deflection using molecular simulations and the CG model, (c) A (5, 5) CNT with periodic boundary conditions, (d) coarse-grained model of the nanotube, and (e) the variation of potential energy versus longitudinal deflection of the tube using molecular and coarse-grained simulations.

The findings reveal an excellent agreement between results obtained from the CG model and molecular simulations with a percentage difference less than 0.5%. It can be concluded that the force field parameters determined for C beads enable to precisely estimate the mechanical behavior of CNTs.

*3.2. Mechanical properties carbon nanotube/polymer composites*

Following the successful examinations of the CG model in predicting the mechanical behavior of PMMA polymer chains and CNTs, we extend the CG model to measure the mechanical properties of CNT/PMMA composites. A simulation unit cell with a size of 5×5×5 nm³ and periodic boundary conditions that contains a (5, 5) CNT bundle is initially constructed. A number of the PMMA chains composed of 30 monomer units are packed into the cubic lattice with a predefined mass density of 1 $g/cc$. The generation is fulfilled by the Amorphous Cell Packing task in Accelrys Materials Studio 7.0. The module builds molecules in a cell with a Monte Carlo fashion, by minimizing close contacts between atoms, whilst ensuring a realistic distribution of torsion angles given for the COMPASS force field. An atomistic RVE of the polymer matrix reinforced by a three-CNT bundle illustrated in Figs. 8 (a) and (b) consists of 11144 atoms. The atomistic system is then mapped onto its own CG model as shown in Figs. 9 (a) and (b). The CG RVE consists of 684 beads, indicating a 16.29 fold decrease in the number of DOF with respect to the full atomistic system.

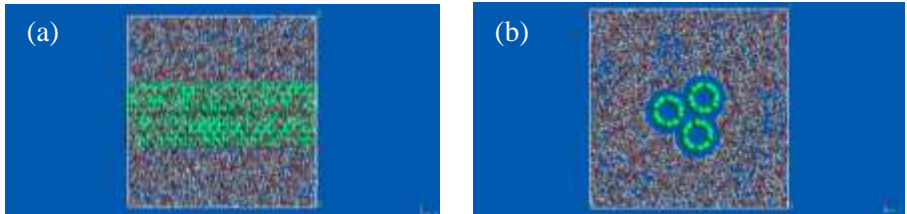

Fig. 8. Atomistic RVE of PMMA polymer matrix with a size of 5×5×5 nm$^3$ reinforced by a nanotube rope made of three (5, 5) SWCNTs: (a) side view, (b) top view. The RVE consists of 11144 atoms.

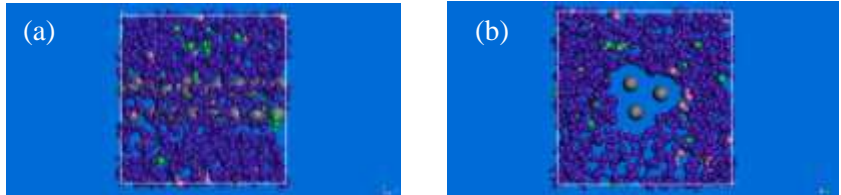

Fig. 9. CG RVE of PMMA polymer matrix with a size of 5×5×5 nm$^3$ reinforced by a nanotube rope made of three (5, 5) SWCNTs: (a) side view, (b) top view. The RVE consists of 684 beads.

In order to find a global minimum energy configuration, a geometry optimization with the convergence criteria of 0.00001 kcal/mol is initially performed to minimize the total energy of the system. Once the minimization process is completed, the system is allowed to equilibrate over the isothermal–isobaric ensemble (NPT) ensemble at room temperature of 298 K and atmospheric pressure of 101 kPa for 2 ns. In the NPT simulations, the time step is set to be 1 and 10 fs for MD and GC simulations, respectively. The Andersen feedback thermostat [39] and the Berendsen barostat algorithm [40] are respectively used for the system temperature and pressure conversions. The mass density versus time of the CNT/polymer composite obtained by MD simulation and the CG model is presented in Fig. 10. From Fig. 10, the mass densities of the system predicted by MD simulation and the CG model are respectively converged to 1.023 and 1.021 g/cc, indicating a percentage difference less than 0.2%. The simulation results confirm the applicability of the defined CG force field parameters in accurately predicting the mass density of CNT/PMMA composites. The NPT simulation is followed by a further energy minimization. The process removes internal stresses in the composite material.

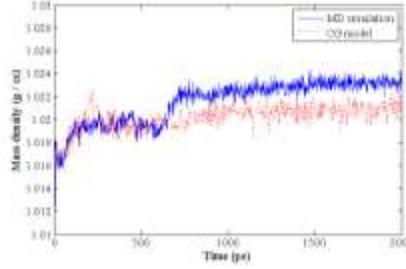

Fig. 10. Mass density versus time of the PMMA polymer matrix reinforced by a CNT bundle made of three (5, 5) nanotube.

After the preparation of the CNT/PMMA composite material, the constant-strain minimization method is applied to the equilibrated system to measure the material properties of the composite. A small strain of 0.01% is applied to the periodic structure in the longitudinal direction (*x*-direction). The application of the tensile strain is accomplished by uniformly expanding the dimensions of the simulation cell in the direction of the deformation and re-scaling the new coordinates of the atoms to fit within the new dimensions. After each increment of the applied strain, the potential energy of the structure is re-minimized keeping the lattice parameters fixed. The total potential energy and the interaction energy are then measured in the minimized structure. This process is repeated for a series of strains. Finally, the variation of the measured potential energies versus applied strain is used to calculate the effective Young's moduli of the interfacial region and composite as $E = \frac{1}{V}\left(\frac{\partial^2 U}{\partial \varepsilon^2}\right)$, where $U$ is the potential energy, $V$ is the unit volume and $\varepsilon$ is strain. Fig. 11 presents the variation of potential energy versus strain of the CNT/PMMA composite illustrated in Figs. 8 and 9 from which the Young's modulus in the axial direction of the composite predicted by molecular simulations and the CG model is calculated to be 72.54 and 73.39 GPa, respectively. The simulation results reveal that the CG model is able to precisely estimate the mechanical properties of CNT/PMMA composites with respect to molecular simulation results as a benchmark.

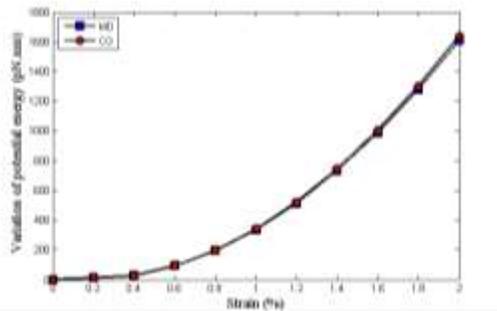

Fig. 11. Potential energy of a PMMA polymer matrix reinforced by a bundle of three CNTs versus strain obtained by molecular simulation and coarse-grained model.

To further explore the applicability of the CG model, the effect of the volume fraction of CNT bundle reinforcements on the mechanical properties of CNT/PMMA composites is presented in Table 1. In simulations of PMMA matrix reinforced by CNT bundles with an infinite length, a RVE with a size of 5×5×5 nm³ and periodic boundary conditions is considered. The predefined mass density of CNT/PMMA composites is set to be 1 $g/cc$. The diameter of the (5, 5) CNT reinforcements is 0.68 nm and their volume fraction varies from 10 to 20 %. To adjust the value of volume fraction of the reinforcements, the number of CNTs in a bundle differs from 3 to 6. Since the polymer matrix does not penetrate the CNTs, the effective volume fraction of a CNT bundle, $c_f$, including cross-sectional area of nanotubes is defined as $c_f = \frac{\pi \left(R_{CNT} + \frac{h_{vdW}}{2}\right)^2}{A_{RVE}}$, where $R_{CNT}$ is the radius of nanotubes, $h_{vdW}$ is the equilibrium van der Waals separation distance between nanotubes and the matrix, and $A_{RVE}$ is the cross-sectional area of the unit cell perpendicular to the CNT bundle axis. The van der Waals separation distance is generally a function of interfacial interactions between nanotubes and the matrix, which is measured to be 0.176 nm in the present simulations. Based on the definition, the volume fraction of three-CNT and six-CNT bundles is obtained to be 10 and 20 %, respectively.

From Table 1, the Young's modulus of CNT/PMMA composite predicted by the CG model increases from 73.39 GPa to 97.13 and 122.05 GPa with an increase in the volume fraction of the CNT bundles from 10 % to 13.38 and 16.73 %. The measured Young's moduli for the PMMA matrix composite reinforced by CNT bundles are in excellent agreement with values of 72.54, 96.05 and 121.30 GPa obtained from molecular simulations, where the fracture volume of nanotubes are 10, 13.38 and 16.73 %, respectively. The simulation results show a percentage increase of 32 and 67% in the stiffness of the CNT/PMMA composite with an increase in the volume fraction of the CNT bundles from 10 % to 13.38 and 16.73%. In addition, the Young's modulus of the CNT/PMMA composite with a CNT bundle volume fraction of 20% increases to 145.49 GPa, which is more than 50 times stiffer than a PMMA polymer material.

Table 1. Mechanical properties of the PMMA polymer matrix with a size of $5 \times 5 \times 5\ nm$ reinforced by a bundle of (5, 5) CNTs. The Young's modulus of PMMA polymer is 2.86 GPa.

| Volume fraction of CNTs (%) | CG model (GPa) | MD (GPa) |
|---|---|---|
| 10 (3 CNTs) | 73.39 | 72.54 |
| 13.38 (4 CNTs) | 97.13 | 96.05 |
| 16.73 (5 CNTs) | 122.05 | 121.30 |
| 20 (6 CNTs) | 145.49 | 145.34 |

After determining the ability of the CG method in modeling CNT/PMMA composites, the potential of the method in predicting the elastic properties of a relatively large RVE is investigated. In simulations, a polymer matrix with a size of 20×20×20 nm³ reinforced by randomly distributed (5, 5) CNTs is considered as illustrated in Fig. 12 (a). The size of the RVE is 64 times greater that the full atomistic unit cell illustrated in Fig. 8. The volume fraction of the nanotubes is set to be 8%, while their lengths vary from 5 to 10 nm. Fig. 12 (b)-(d) shows the random distribution of the CNTs in the polymer matrix. The PMMA chains are also packed into the cubic lattice with a predefined mass density of 1 $g/cc$. The CG RVE consists of 42890 beads, which is equivalent to a full atomistic system with 679750 atoms. The CG model enables a significant decrease in the number of DOF of the system as high as 15.85 times compared to the full atomistic system. From simulation results presented in Table 2, the average Young's modulus of the PMMA polymer matrix is obtained to $E_{ave} = 2.87\ GPa$, which is in an excellent agreement with MD simulation results of 2.86 GPa reported in Ref. [8]. The average Young's modulus of the PMMA polymer composite reinforced by (5, 5) CNTs with length of 5 nm increases to 3.70 GPa as presented in Table 2, revealing a percentage increase of 29% in the stiffness of the PMMA polymer owing to the application of the (5, 5) CNT reinforcement. The simulation results show a percentage increase of 56% in the stiffness of the CNT/PMMA composite with an increase in the length of CNT reinforcements from 5 to 10 nm.

Table 2. Effect of length of CNT reinforcements on the mechanical properties of a PMMA polymer matrix with a size of $20 \times 20 \times 20\ nm$. The volume fraction of nanotubes is set to be 10%.

| Length (nm) | $E_1$ (GPa) | $E_2$ (GPa) | $E_3$ (GPa) | $E_{ave}$ (GPa) | Percentage increase (%) |
|---|---|---|---|---|---|
| Pure polymer | 2.85 | 2.88 | 2.87 | 2.87 | - |
| 5 | 3.75 | 3.61 | 3.74 | 3.70 | 29 |
| 10 | 4.42 | 4.50 | 4.49 | 4.47 | 56 |

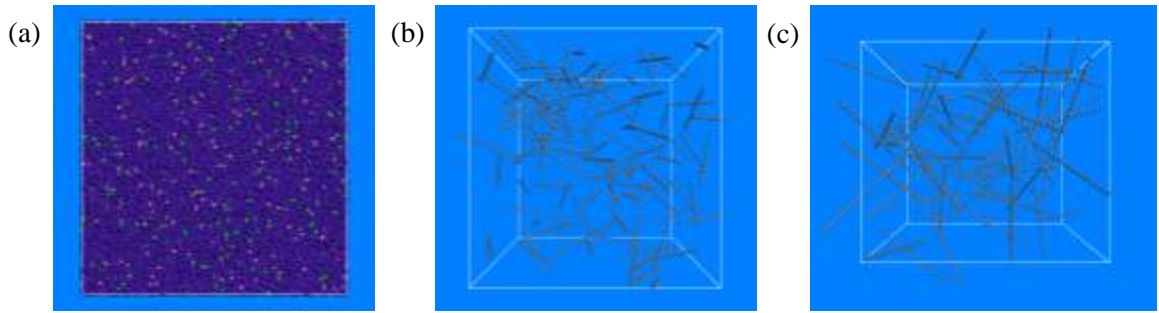

Fig. 12. (a) A (5, 5) CNT with periodic boundary conditions, (b) coarse-grained model of the nanotube, and (c) the variation of potential energy versus longitudinal deflection of the tube using molecular and coarse-grained simulations.

## 4. Conclusions

A CG model is developed for studying the mechanical behavior of PMMA polymer matrix reinforced by nanotube bundles made of (5, 5) SWCNTs. A distinguishing feature of the CG model is that it captures the non-bonded interactions between the nanotube reinforcements and polymer chains. The feature in conjunction with a significant decrease in the number of DOF allows studying the material properties of the reinforced polymer composites in a wide range of length and time scales compared to a full atomistic description. The CG potentials are calibrated using results from full atomistic simulations. The model enables a significant decrease in the number of DOF of a CNT/PMMA RVE with a size of 5×5×5 nm$^3$ up to 16.29 times compared to full atomistic simulations. The Young's moduli of the nanocomposites predicted by the CG model are verified with molecular simulation results, revealing an excellent agreement between results obtained from the CG model and MD simulations. The effect of the volume fraction of CNT bundle reinforcements on the elastic properties of the polymer matrix nanocomposites is investigated. The CG results demonstrate that the Young's modulus of a CNT/PMMA composite with a nanotube bundle volume fraction of 20% significantly increases to 145.49 GPa, which is more than 50 times stiffer than a pure PMMA polymer material. The application of the CG model in simulating large unit cells with a size of 20×20×20 nm$^3$ and randomly distributed CNTs promises the applicability of the model across a wide range of length scales from nano to mesoscale. In the future, the CG model can also be utilized and developed to predict the mechanical properties of polymer composites reinforced by functionalized SWCNs and multi-walled CNTs subjected to static and dynamic loading.

**Acknowledgments**

The authors thank the support of the European Research Council-Consolidator Grant (ERC-CoG) under grant "Computational Modeling and Design of Lithium-ion Batteries (COMBAT)".

**Figure Captions**

Fig. 1. (a) Two monomers of a PMMA polymer chain, (b) CG model of the monomers made of two P beads, and (c) the variation of potential energy versus deformation using molecular simulations.

Fig. 2. (a) A (5, 5) CNT with 10 rings of carbon atoms, where each 5 rings is modeled as a bead in coarse-grained model, (b) CG model of the CNT made of two C beads, and (c) the variation of potential energy versus deformation using molecular simulations.

Fig. 3. (a) Three monomers of a PMMA polymer chain, (b) the CG model of the three-monomer system made of three P beads, and (c) the variation of potential energy versus the bond angle ($\theta$) using molecular simulations.

Fig. 4. (a) A (5 ,5) CNT with fifteen atomic rings, (b) the CG model of the nanotube made of three C beads, and (d) the variation of potential energy versus bond angle of the CNT using molecular simulations.

Fig. 5. (a) Four monomers of a PMMA polymer chain, (b) the CG model of the four-monomer system made of four P beads, and (c) the variation of potential energy versus the dihedral angle ($\phi$) using molecular simulations.

Fig. 6. (a) Two monomers of PMMA polymer, (b) coarse-grained model of the monomers, and (c) coarse-grained model of the monomers, and (c) the variation of vdW energy versus distance between the two monomers using molecular simulations.

Fig. 7. (a) A PMMA polymer chain made of 30 monomers and its CG model made of 30 P beads, (b) the variation of potential energy versus tensile longitudinal deflection using molecular simulations and the CG model, (c) A (5, 5) CNT with periodic boundary conditions, (d) coarse-grained model of the nanotube, and (e) the variation of potential energy versus longitudinal deflection of the tube using molecular and coarse-grained simulations.

Fig. 8. Atomistic RVE of PMMA polymer matrix with a size of 5×5×5 nm$^3$ reinforced by a nanotube rope made of three (5, 5) SWCNTs: (a) side view, (b) top view. The RVE consists of 11144 atoms.

Fig. 9. CG RVE of PMMA polymer matrix with a size of 5×5×5 nm$^3$ reinforced by a nanotube rope made of three (5, 5) SWCNTs: (a) side view, (b) top view. The RVE consists of 684 beads.

Fig. 10. Mass density versus time of the PMMA polymer matrix reinforced by a CNT bundle made of three (5, 5) nanotube.

Fig. 11. Potential energy of a PMMA polymer matrix reinforced by a bundle of three CNTs versus strain obtained by molecular simulation and coarse-grained model.

Fig. 12. (a) A (5, 5) CNT with periodic boundary conditions, (b) coarse-grained model of the nanotube, and (c) the variation of potential energy versus longitudinal deflection of the tube using molecular and coarse-grained simulations.

**Table Captions**

Table 1. Mechanical properties of the PMMA polymer matrix with a size of $5 \times 5 \times 5\ nm$ reinforced by a bundle of (5, 5) CNTs. The Young's modulus of PMMA polymer is 2.86 GPa.

Table 2. Effect of length of CNT reinforcements on the mechanical properties of a PMMA polymer matrix with a size of $20 \times 20 \times 20\ nm$. The volume fraction of nanotubes is set to be 10%.